# Performance of MC-MC CDMA Systems with Nonlinear Models of HPA

Labib Francis Gergis

Misr Academy for Engineering and Technology, Mansoura, Egypt

IACSIT Senior Member, IAENG Member

*drlabeeb@yahoo.com*

## Abstract

*A new wireless communication system denoted as Multi-Code Multi-Carrier CDMA (MC-MC CDMA), which is the combination of Multi-Code CDMA and Multi-Carrier CDMA, is analyzed in this paper. This system can satisfy multi-rate services using multi-code schemes and muti-carrier services used for high rate transmission. The system is evaluated using Traveling Wave Tube Amplifier (TWTA). This type of amplifiers continue to offer the best microwave high power amplifiers (HPA) performance in terms of power efficiency, size and cost, but lag behind Solid State Power Amplifiers (SSPA's) in linearity. This paper presents a technique for improving TWTA linearity. The use of pre-distorter (PD) linearization technique is described to provide TWTA performance comparable or superior to conventional SSPA's. The characteristics of the PD scheme is derived based on the extension of Saleh's model for HPA.*

## Keywords

*Multi-Carrier CDMA, Multi-Code CDMA, MC-MC CDMA, TWTA, HPA, AM/AM, AM/PM, Saleh Model.*

## 1. Introduction

Multi-Code CDMA and Multi-Carrier CDMA have attracted a lot of attention from researchers due to their perceived high rate transmission capability. In Multi-Code CDMA, researchers have investigated the systems performance in different fading channel [1] and suggested many schemes to improve the performance [2]. In Multi-Code CDMA, the input data streams are first split into several substreams in parallel and then orthogonal codes are multiplied for each substream.

In Multi-Carrier CDMA [3], the input data streams are first split into several substreams in parallel, like in Multi-Code CDMA and then modulates several subcarriers with each substream before transmitting the signals. Similarly with Multi-Code CDMA, Multi-Carrier CDMA is analyzed with different fading channels, and researchers have suggested schemes to improve the system performance [4], and [5].

In [6], [7], [8], and [9] Multi-Code Multi-Carrier CDMA system was evaluated and compared with both single code multi-carrier CDMA system and multi-code CDMA system with single carrier in a frequency selective fading channel.

Power amplifiers (PA's) are vital components in many communication systems. The linearity of a PA response constitutes an important factor that ensures signal integrity and reliable performance of the communication system. High power amplifier (HPA) in microwave range suffers from the effects of amplitude modulation to amplitude modulation distortion (AM/AM), and amplitude modulation to phase modulation distortion (AM/PM), during conversions caused by the HPA amplifier. These distortions can cause intermodulation (IM) distortion, which is undesirable to system designs. The effects of AM/AM and AM/PM distortions can cause the bit error rate performance of a communication channel to be increased.

The amplitude and phase modulation distortions are minimized using linearization methods. The linearization method requires modeling the characteristics of the amplitude distortion and phase distortion of the HPA. A Saleh model has been used to provide the linearization method and applied to measured data from HPA that characterize the distortion caused by the HPA. The measured data provides a performance curve indicating nonlinear distortion. The forward Saleh model is a math equation that describes the amplitude and phase modulation distortions of the HPA [10], and [11]. The amount of desired pre-distortion (PD) linearization is then determined to inversely match the amount of distortion for canceling out the distortion of the HPA.

The remainder of the paper is organized as follows. Section 2 discusses the proposed MC-MC CDMA system model. The bit error rate performance analysis of the proposed system is derived in Section 3. In section 4, numerical results were presented, showing an improvement in the performance of MC-MC CDMA transmission system over the nonlinear channels. Section 5, summarized the conclusions had obtained through this paper.

## 2. System Model

### 2.1 Transmitter Model

The transmitter of the system model shown in Figure 1, is made up of two parts: the multi-code part and the multi-carrier part.

The multi-code part converts serial input data streams into parallel substreams, spreads each parallel substream to produce code division multiplexed bits with multi-codes, which are orthogonal to each other. Then, all substreams are summated to produce super-stream, $B_k(t)$.

In multi-carrier part, the super stream is serial-to-parallel (S/P) converted again, spread with a user specified Pseudo-random noise (PN) sequence, and modulated with orthogonal multi-carriers. Finally, the signal which is summated in the multi-carrier part is transmitted by the transmitter.

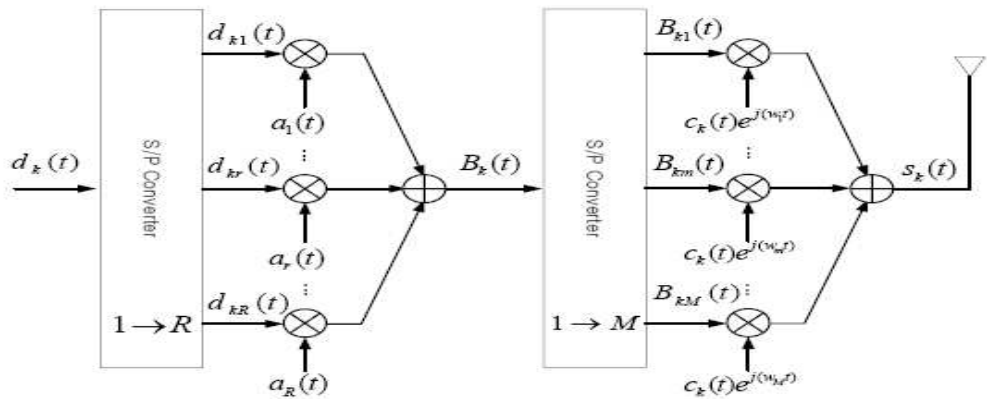

Figure 1. Transmitter Structure for the MC-MC CDMA System

By considering the system in which there are K users transmitting information simultaneously in cellular system. The transmitter for the $k^{th}$ using BPSK modulation, is shown in Figure 1. The data signal $d_k(t)$ is a random complex sequence representing the data bit and having a symbol rate of ($RM/T$), where $R$ is the number of substreams, $M$ is the number of sucarriers in the multi-code, and $T$ is the bit duration per input data stream, $d_k(t)$ can be expressed as

$$d_k(t) = d_k \; P_x(t) \tag{1}$$

where $x = T/RM$

After S/P conversion of the input data stream and modulating with orthogonal codes, the resulting output is given by

$$B_k(t) = \sum_{r=1}^{R} d_{kr}(t) \; a_r(t) \tag{2}$$

where $d_{kr}(t)$ is a substream S/P converted from the input data stream and is given by

$$d_{kr}(t) = d_{kr} \; P_{T/m}(t) \tag{3}$$

and $a_r$ is an orthogonal code set for the $r^{th}$ substream of the $R$ substreams, defined by

$$a_r(t) = a^n{}_r \; P_{TNa}(\,t - n\,T_{Na}\,) \tag{4}$$

where $T_{Na}$ is the chip duration, $N_a$ is the code length of the orthogonal codes, $a^n{}_r$ is the $n^{th}$ value $\in \{\pm 1\}$ of the code $a_r$.

The transmitted signal $S_k(t)$ as a result of the summation of parallel super-substreams that S/P converted from the super-stream, multiplied by each subcarrier, and spread by the PN sequence. This is given by [6]

$$S_k(t) = \sqrt{2\,P_k} \; \sum_{m=1}^{M} Re\{\,B_{km}(t) \; C_k(t) \; e^{j(wmt)}\,\} \tag{5}$$

$$S_k(t) = \sqrt{2\,P_k} \; \sum_{m=1}^{M} \sum_{r=1}^{R} Re\{\,d_{krm}(t) \; a_r(t) \; C_k(t) \; e^{j(wmt)}\,\} \tag{6}$$

where

a) $P_k$ is the signal power of user $k$ distributed among the carriers, assuming $P_1 = P_2 = \ldots = P_k = P$.

b) $B_{km}(t)$ is the $m^{th}$ super-substream S/P converted from super-stream $B_k(t)$ with a bit rate of ($R/T$). After S/P conversion, the symbol duration increases $M$ times.

c) $d_{krm}(t)$ is the data symbol of $r^{th}$ substream of the $m^{th}$ super-stream with value

$$d_{krm}(t) = d_{krm} \; P_T(t) \tag{7}$$

*d)* $C_k(t)$ is the PN sequence defined as

$$C_k(t) = \sum_{s=0}^{N_c-1} C^s_k \ P_{TNc} (t - s \ T_{Nc}) \tag{8}$$

$$T_{Nc} = T / N_c \tag{9}$$

$N_c$ is the length of a PN sequence and $C^s_k(t)$ is the $s^{th}$ value of the PN sequence.
The subcarrier $e^{j(w_m)}$ is the mth subcarrier having frequency $f_m$, defined by

$$w_m = 2\pi f_m, f_m = R_m / T, \text{ and } m = 1, 2, \ldots, M. \tag{10}$$

### 2.2 Nonlinear Model

High power amplifiers exist in almost all wireless communication links. Due to various nonlinear electronic components inside them, these power amplifiers are nonlinear devices. Power amplifiers exhibit nonlinear distortion in both amplitude and phase. The amplitude conversion is referred as Amplitude Modulation to Amplitude Modulation (AM/AM) conversion, and phase conversion is referred as Amplitude Modulation to Phase Modulation (AM/PM) conversion.

A major type of power amplifiers typically used in communication systems, is travelling wave tube amplifiers (TWTA).
The classical and most often used nonlinear model of power amplifier is Saleh's model [10]. It is a pure nonlinear model without memory. The output of HPA defined in Figure .2, is expressed as

$$S_y = A [U_x] e^{j(\alpha x + \Phi[Ux])} \tag{11}$$

where the input-output functional relation of the HPA has been defined as a *transfer function*.
The equations define this base-band model of HPA as two modulus dependent *transfer functions* are defined as [9] and [10] :

$$A[U_x] = \alpha_a \ U_x \ / \ 1 + \beta_a \ U^2_x$$
$$\Phi[U_x] = \alpha_\Phi \ U_x \ / \ 1 + \beta_\Phi \ U^2_x \tag{12}$$

where $A[U_x]$ and $\Phi[U_x]$ are the corresponding AM/AM and AM/PM characteristics respectively, both dependent exclusively on $U_x$, which is the input modulus to HPA.
The values of $\alpha_a, \beta_a, \alpha_\Phi$ and $\beta_\Phi$ are defined in [10]. The corresponding AM/AM and AM/PM curves so scaled are depicted graphically in Figure 2.

The HPA operation in the region of its nonlinear characteristic causes a nonlinear distortion of a transmitted signal, that subsequently results in increasing the bit error rate (*BER*), and the out-of-band energy radiation ( spectral spreading ).
The operating point of HPA is defined by input back-off (*IBO*)$_{dB}$ parameter which corresponds to the ratio of saturated input power ($P_{maxin}$), and the average input power ( $P_x$ ), is defined as [11] :

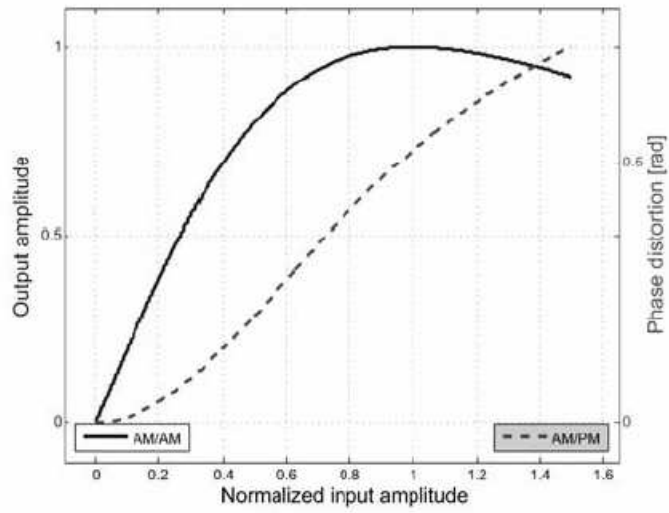

**Figure 2. AM/AM and AM/PM characteristics of the normalized Saleh model**

$$IBO_{dB} = 10 \log_{10} (P_{max\,in} / \overline{P_x}) \qquad (13)$$

and out back-off (*OBO*) $_{dB}$ corresponds to the ratio between the saturated and average output power($P_{maxout}$), and the average input power ($\overline{P_y}$), and is defined as :

$$OBO_{dB} = 10 \log_{10} (P_{max\,out} / \overline{P_y}) \qquad (14)$$

The measure of effects due to the nonlinear HPA could be decreased by the selection of relatively high values of *IBO*.

The graphical representation of these two parameters is shown in Fig. 3, with dB units.

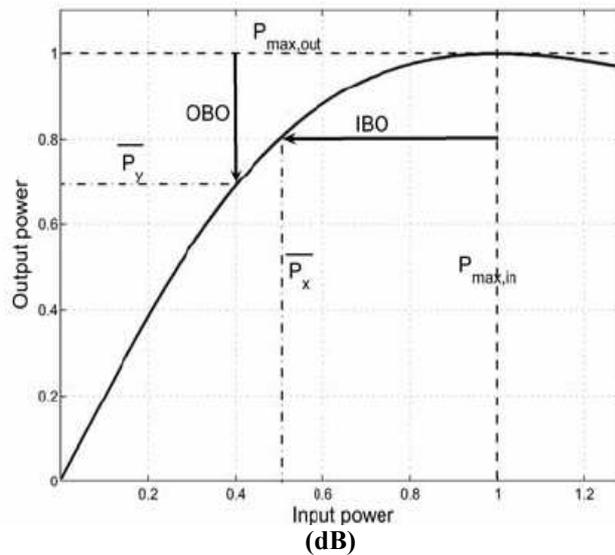

**Figure. 3 Graphical representation of IBO and OBO.**

### 2.3 Receiver Model

The receiver structure is shown in Figure 4, where signals consisting of *M* carriers are demodulated by a locally generated carriers, despread by a user specific PN sequence, and parallel-to-serial (P/S) converted to produce $\hat{B}_k(t)$. Then $B_k(t)$ is despread with $a_r$ code sequence, correlated over one symbol period then P/S converted to recover $\hat{d}_{kr}$.

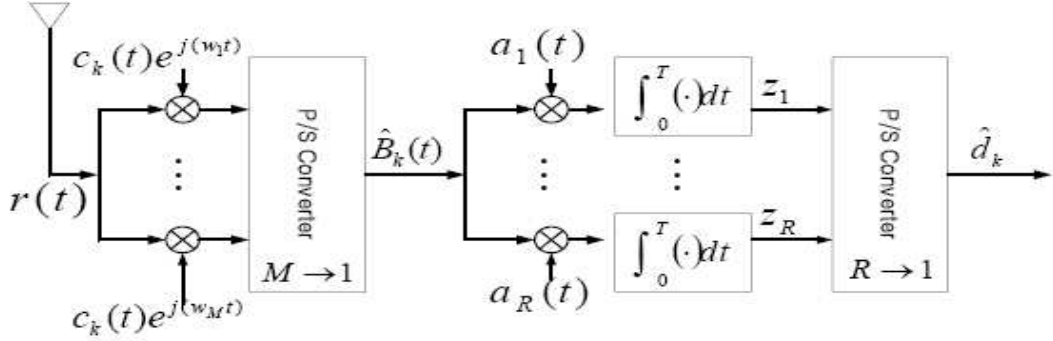

**Figure 4. Receiver Structure for the MC-MC CDMA System**

The overall received signal equation corrupted with Additive White Gaussian Noise (AWGN) is given by [8]:

$$r(t) = \sum_{k=1}^{k} y_k(t) + n(t)$$

$$= \sqrt{2P} \; A_{11} \, Re \left[ d_{111}(t-\tau_{11}) \, a_1(t-\tau_{11}) c_1(t-\tau_{11}) \, e^{j\{w_1(t-\tau_{11})+\Phi_{11}\}} \right]$$

$$+ \sqrt{2P} \sum_{l=2}^{L} A_{1l} \, Re \left[ d_{111}(t-\tau_{1l}) \, a_1(t-\tau_{1l}) c_1(t-\tau_{1l}) \, e^{j\{w_1(t-\tau_{1l})+\Phi_{1l}\}} \right]$$

$$+ \sqrt{2P} \sum_{r=2}^{R} \sum_{l=1}^{L} A_{1l} \, Re \left[ d_{1r1}(t-\tau_{1l}) \, a_r(t-\tau_{1l}) c_1(t-\tau_{1l}) \, e^{j\{w_1(t-\tau_{1l})+\Phi_{1l}\}} \right]$$

$$+ \sqrt{2P} \sum_{m=2}^{M} \sum_{r=1}^{R} \sum_{l=1}^{L} A_{1l} \, Re \left[ d_{1rm}(t-\tau_{1l}) \, a_r(t-\tau_{1l}) c_1(t-\tau_{1l}) \, e^{j\{w_m(t-\tau_{1l})+\Phi_{1l}\}} \right]$$

$$+ \sqrt{2P} \sum_{K=2}^{K} \sum_{m=1}^{M} \sum_{r=1}^{R} \sum_{l=1}^{L} A_{kl} \, Re \left[ d_{krm}(t-\tau_{1l}) \, a_r(t-\tau_{kl}) c_k(t-\tau_{kl}) \right.$$

$$\left. \cdot e^{j\{w_m(t-\tau_{kl})+\Phi_{kl}\}} \right]$$

$$+ n(t) \tag{15}$$

The received signal can be expressed in six components as[8] :

$$r(t) = r_{DS}(t) + r_{MPI}(t) + r_{ISSI}(t) + r_{ICI}(t) + r_{MUI}(t) + n(t) \qquad (16)$$

where
  a) $r_{DS}(t)$ is the desired signal, corresponding to reference user's ($k=1$), reference substream ($r=1$), reference carrier ($m=1$), and reference path ($l=1$).
  b) $r_{MPI}(t)$ is the interference caused by the propagation of the desired signal and corresponds to reference user's ($k=1$), reference substream ($r=1$), and reference carrier ($m=1$), from other paths ($l \neq 1$), which is called Multipath Interference (MPI)
  c) $r_{ISSI}(t)$ is the interference caused by other substreams except the reference substream, which is called Inter-Substream Interference (ISSI)
  $r_{ICI}(t)$ is the interference caused by other carriers except the reference carrier, which is called Inter-Carrier Interference (ICI)
  d) $r_{MUI}(t)$ is the interference caused by other users except the reference user ($k=1$), which is commonly known as Multi User Interference (MUI)
  e) $n(t)$ is the AWGN component.

At the receiver, the received signal is first demodulated by locally generated carrier and then despread by a user specific code sequence before P/S conversion.
The output signals of the P/S is despread again by each orthogonal code for multicode component in order to recover a substream before correlating over a period $T$. Finally R substreams are recovered from the correlated outputs.

The output of the correlator also may be decomposed into six components [12]:

$$z_1 = z_{DS} + z_{MPI} + z_{ISSI} + z_{ICI} + z_{MUI} + z_n \qquad (17)$$
$$z_1 = z_{DS} + I_{Total} \qquad (18)$$

where each component is similarly defined as in (12), and $z_n$ is the correlated AWGN component, and $I_{Total}$ is the sum of the all interference terms and AWGN component.
The desired signal power can be defined from the first term of (13). Using $(d^I_{111})^2 = 1$

$$(z_{DS})^2 = S = (P/2)(A_{11})^2 T^2 \qquad (19)$$

## 3. Bit Error Rate Performance Analysis ( BER )

The average bit error probability can be expressed as [8] :

$$\overline{P_e} = \int_0^\infty f(A_{11}) \; P_e(A_{11}) \, dA \qquad (20)$$

where $f(A_{11})$ is the pdf of random variable $A_{11}$.

$$P_e(A_{11}) = 1/2 \; erfc(\gamma) \qquad (21)$$

where $erfc(.)$ is the complementary error function, and $\gamma$ is defined as [9] :

$$\gamma = S / \sigma^2_{Total} = (P/2)(A_{11})^2 T^2 / \sigma^2_{Total} \qquad (22)$$

$\sigma^2_{Total}$ is the total variance for the all interference terms and AWGN component, it can be written as

$$\sigma^2_{Total} = \sigma^2_{MPI} + \sigma^2_{ISSI} + \sigma^2_{ICI} + \sigma^2_{MUI} + \sigma^2_n \qquad (23)$$

## 4. Results and Discussions

In this section, the BER performance of MC-MC CDMA is derived and compared with Multi-Code CDMA system, and Multi-Carrier CDMA system, and some properties of MC-MC CDMA are observed. Fig. 5, shows the comparison assuming that all systems use BPSK modulation technique, number of substreams for Multi-Code CDMA and MC-MC CDMA (*R*) equals 8, number of carriers for Multi-Carrier CDMA and MC-MC CDMA (M) equals 8, and number of users (K) equals 20. The Figure indicates that the MC-MC CDMA system has the lowest BER performance of all the system compared. That means that MC-MC CDMA system uses several times more bandwidth compared to MC - CDMA with the same data rate.

Due to the gain which comes from orthogonality between code sequences and frequency spreading gain, MC-MC CDMA system shows better performance than MC-CDMA [8].

Fig. 6, illustrates the BER performance of MC-MC CDMA *(R=M=8)* as a function of number of users *(K = 1, 10, and 50)*. It is shown that with increasing simultaneous users, BER increase, and the performance decreases.

Fig. 7 shows the effect of the number of subcarriers M on the BER performance of MC-MC CDMA system *(R=8, K=20)* . It is clear that by increasing M values, a higher BER performance is gained.

The BER performance of MC-MC CDMA system *(R=M=8, K=20)* is expressed under the case of nonlinearity distortions (AM/AM and AM/PM), for different values of *IBO* (*IBO* = 7 dB and *IBO* = 9 dB), compared with the case of no nonlinear ( that had linearized by PD), is illustrated in Fig. 8. It is shown that it is highly recommended to use linearization devices at the transmitter side in order to suppress the undesirable nonlinearity effects and to get improved bit error performance.

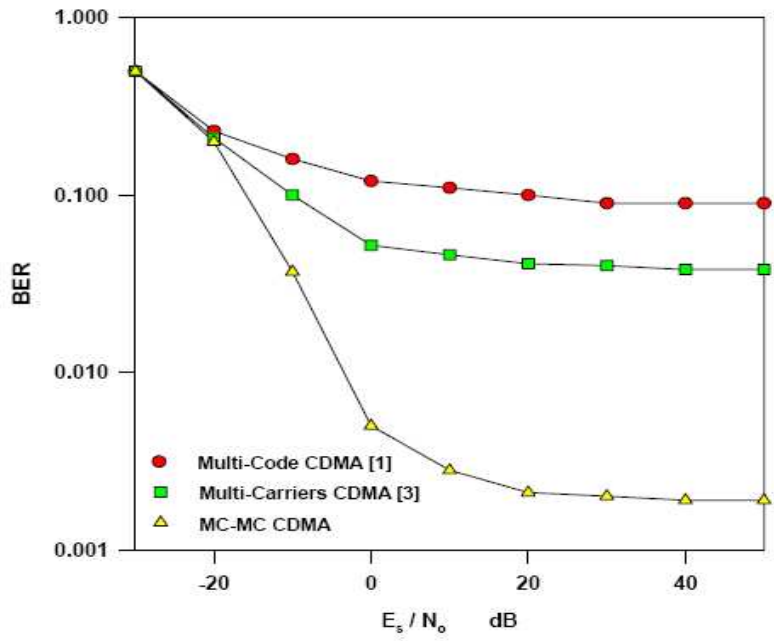

**Figure 5. Bit Error Rate Respect to Multi-Code CDMA, Multi-Carriers CDMA and MC-MC CDMA System**

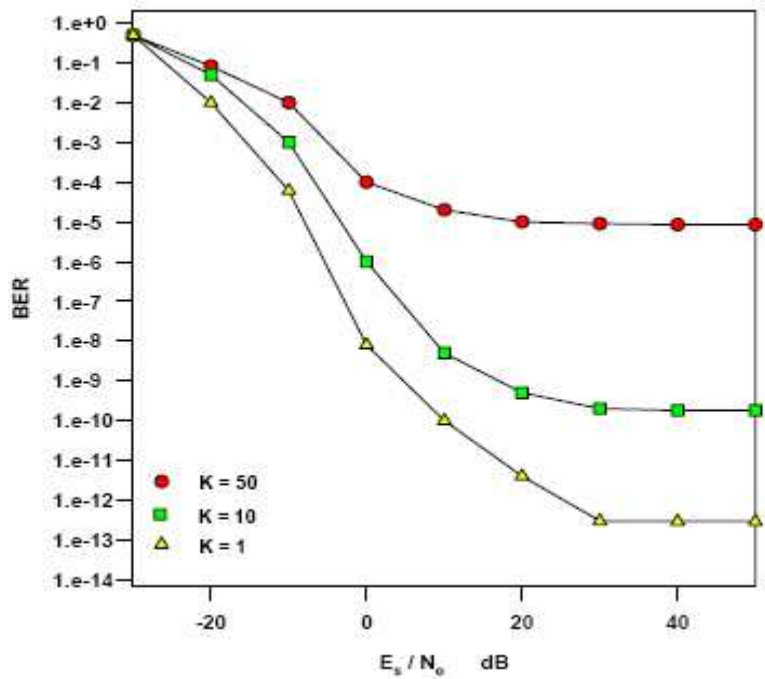

**Figure 6. BER Performance of MC-MC CDMA System as a Function of Number of Users ( K )**

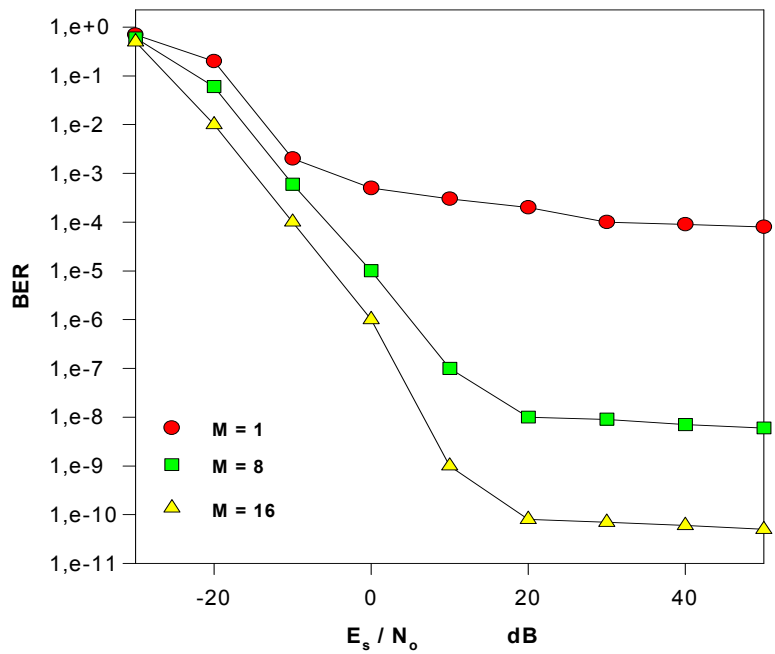

**Figure 7. BER Performance of MC-MC CDMA System as a Function of Number of Subcarriers ( M )**

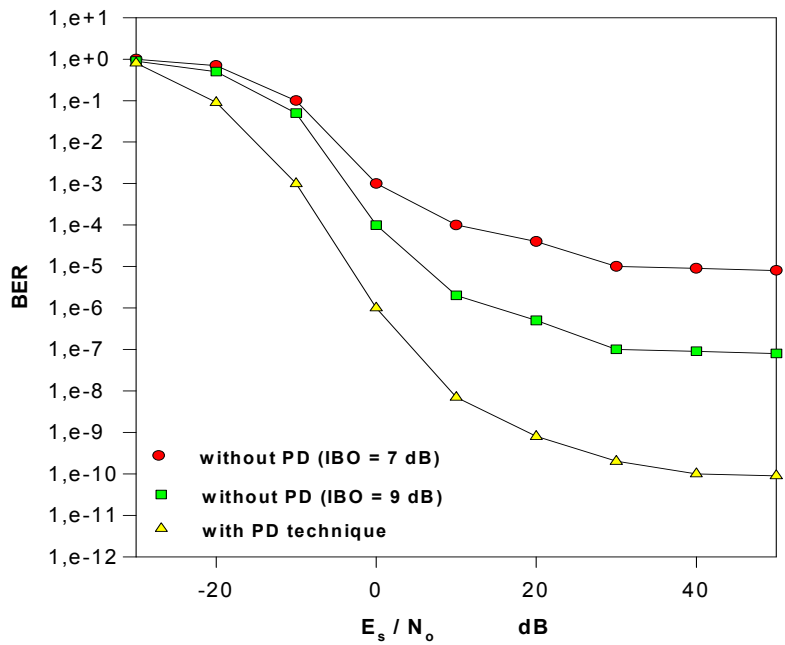

**Figure 8. Effect of Using Pre-Distorter Technique ( PD ) on the BER Performance of MC-MC CDMA System**

## 5. CONCLUSIONS

In this paper, the analysis and performance results of MC-MC CDMA system is presented in terms of BER. The performance of the MC-MC CDMA is compared to the performance of Multi-Carrier CDMA, and Multi-Code CDMA. It was shown that the MC-MC CDMA system outperformed other systems.

The effects of nonlinearities (AM/AM and AM/PM) were analyzed. It is shown that these effects can be compensated by using PD technique. From analytical results, it is confirmed that PD system with MC-MC CDMA gives a good performance improvements compared to MC-MC CDMA signals without PD.